\documentclass[12pt]{article}
\usepackage{epsfig,amsmath,amssymb,mathrsfs}

  {\end{list}}%

\makeatletter
\renewcommand{\section}{\setcounter{equation}{0}\@startsection
 {section}%
 {1}%
 {0pt}%
 {-1\baselineskip}%
 {0.4\baselineskip}%
 {\bfseries\large}}%
\renewcommand{\subsection}{\@startsection
 {subsection}%
 {2}%
 {0pt}%
 {-0.75\baselineskip}%
 {0.2\baselineskip}%
 {\bfseries}}%
\renewcommand{\subsubsection}{\@startsection
 {subsubsection}%
 {3}%
 {0pt}%
 {-0.5\baselineskip}%
 {0.1\baselineskip}%
 {\sc}}%
\makeatother

\DeclareMathAlphabet{\mathpzc}{OT1}{pzc}{m}{it}

\def\AA{{\mathbb A}}
\def\MM{{\mathbb M}}

\def\g5{\gamma_{5}}

\def\idp{\int\!\! \frac{d^4\!p}{(2\pi)^4}}
\def\idpp{\int\!\! \frac{d^4\!p_1}{(2\pi)^4}\frac{d^4\!p_2}{(2\pi)^4}}
\def\idppp{\int\!\! \frac{d^4\!p_1}{(2\pi)^4}\frac{d^4\!p_2}{(2\pi)^4}\frac{d^4\!p_3}{(2\pi)^4}}
\def\idpppp{\int\!\! \frac{d^4\!p_1}{(2\pi)^4}\frac{d^4\!p_2}{(2\pi)^4}\frac{d^4\!p_3}{(2\pi)^4}\frac{d^4\!p_4}{(2\pi)^4}}
\def\idpq{\int\!\! \frac{d^4\!p}{(2\pi)^4}\frac{d^4\!q}{(2\pi)^4}}

\def\MM{{\rm I\!\!\, M}}

\newcommand{\bea}{\begin{eqnarray}}
\newcommand{\eea}{\end{eqnarray}}
\newcommand{\beann}{\begin{eqnarray*}}
\newcommand{\eeann}{\end{eqnarray*}}
\newcommand{\ba}{\begin{array}}
\newcommand{\ea}{\end{array}}

 \def\g {\gamma}

\begin{document}
 \begin{titlepage}
\rightline{FTI/UCM 137-2012}
\vglue 33pt

\begin{center}

{\Large \bf Computing the $\theta$-exact Seiberg-Witten map for arbitrary gauge groups.}\\
\vskip 1.0true cm
{\rm C. P. Mart\'{\i}n}\footnote{E-mail: carmelo@elbereth.fis.ucm.es}
\vskip 0.1 true cm
{\it Departamento de F\'{\i}sica Te\'orica I,
Facultad de Ciencias F\'{\i}sicas\\
Universidad Complutense de Madrid,
 28040 Madrid, Spain}\\
\vskip 0.85 true cm
{\leftskip=50pt \rightskip=50pt \noindent
We discuss how to obtain $\theta$-exact Seiberg-Witten maps by expanding in the gauge coupling constant or, equivalently, in the number of ordinary gauge fields. We do so for
arbitrary compact gauge groups in arbitrary unitary representations. For gauge  and matter fields, we  fully work out  $\theta$-exact non-hybrid Seiberg-Witten
maps up to order three in the number of ordinary gauge fields.
\par}
\end{center}

\vspace{9pt}
\vfill
\end{titlepage}

\section{Introduction}
A dozen years ago, it was put forward a formalism --the enveloping-algebra formalism-- which can be
used to construct noncommutative gauge theories that are deformations of ordinary theories with
simple gauge groups in arbitrary unitary representations~\cite{Madore:2000en, Jurco:2000ja,
  Jurco:2001rq}. No other formalism is known that does this generic job.
The chief feature of the formalism at hand--see
ref.~\cite{Blaschke:2010kw}, for a review-- is that both
noncommutative gauge fields and infinitesimal noncommutative gauge
transformations take values on the universal enveloping algebra of the
corresponding Lie algebra, and are defined in terms of the ordinary fields by the appropriate Seiberg-Witten maps. The construction  of a
noncommutative generalisation --called the Noncommutative Standard Model-- of
the Standard Model demands the use of the enveloping-algebra formalism  if no
new particles are introduced --for noncommutative generalisations of the Standard Model outside the
enveloping-algebra formalism see refs.~\cite{Chaichian:2001py, Khoze:2004zc, Arai:2006ya}.
The Noncommutative Standard Model was introduced in ref.~\cite{Calmet:2001na}. The formulation  of noncommutative deformations of the ordinary SO(10) and $\text{E}_6$ Grand Unified models also requires, at present, the use
of the enveloping formalism, as has been done in refs.~\cite{Aschieri:2002mc, Martin:2010ng} --see, however, ref.~\cite{Bonora:2000td} for an alternative formulation of  SU(5) noncommutative Grand Unified models.

By considering the Seiberg-Witten map as a formal power series in the noncommutativity matrix parameter $\theta^{\mu\nu}$,  a fair amount of phenomenological consequences --which might be tested against the data from the LHC-- have been
drawn from the noncommutative Standard Model: refs.~\cite{Melic:2005su, Alboteanu:2006hh, Buric:2007qx,
  Tamarit:2008vy, Haghighat:2010up},
to quote only a few --the reader may wish to find further information in
ref.~\cite{Trampetic:2009vy}.
Besides, renormalisability
~\cite{Buric:2005xe, Buric:2006wm, Buric:2007ix, Martin:2009sg, Tamarit:2009iy, Martin:2009vg},
anomaly freedom~\cite{Martin:2002nr, Brandt:2003fx} and  existence of classical solutions~\cite{Martin:2005vr, Martin:2006px, Stern:2008wi} are other issues
which have been studied for noncommutative gauge theories formulated within the
enveloping-algebra formalism by expanding in powers of $\theta^{\mu\nu}$.

Given a noncommutative theory defined by means of enveloping-algebra formalism,
the expansion in powers of $\theta^{\mu\nu}$ of  the Seiberg-Witten map leads to Green's functions that
are power series in  $\theta^{\mu\nu}$. These Green's functions, in turn,
yield physical predictions which  can be truly trusted only if the energy is well below the noncommutativity energy scale
$\Lambda_{\text{NC}}$.
Indeed,  one should be prepared to accept that when the momenta are well above $\Lambda_{\text{NC}}$ the rapidly
oscillating Moyal exponentials --which tell us that the fields are propagating in a noncommutative space-- partially modify --rendering UV finite some otherwise UV divergent contributions-- the UV behaviour of the theory and connect it with the deep IR region. This is of course the famous noncommutative  UV/IR  mixing ~\cite{Minwalla:1999px} which is known to occur in noncommutative $U(N)$ theories and which can be thought as a consequence of these noncommutative theories being the low-energy limit of the appropriate open string theories, where the former also occurs~\cite{Gomis:2000bn, Martin:2000bk}. Of course, this UV/IR mixing cannot be exhibited at any finite order in expansion in  powers of  $\theta^{\mu\nu}$. Hence, to study the UV/IR mixing phenomenon in noncommutative gauge theories
defined within the enveloping-algebra formalism, one has to use expressions for Seiberg-Witten maps that are
exact in $\theta^{\mu\nu}$ --although they may be power series in some other parameter: the coupling constants being
a first choice if the noncommutative Green's functions are to be defined --keeping an eye on phenomenological applications-- by using standard Feynman-diagram technics. It is advisable to stress that the enveloping-algebra
formalism rests on the existence of the Seiberg-Witten map, not on the existence of this map as a power series expansion in  $\theta^{\mu\nu}$.

Let us mention that is has not been shown yet whether noncommutative gauge theories defined by means of the
enveloping-algebra formalism may find accommodation within string theory --F-theory~\cite{Cecotti:2009zf} may help us in doing so-- and that our experience with noncommutative $U(N)$ gauge theories shows that the UV/IR mixing phenomenon has to be dealt with if these noncommutative are to have a stringy derivation.

The authors of ref.~\cite{Schupp:2008fs} were the first to show, in the $U(1)$ case with fermions in the adjoint, that if the $\theta^{\mu\nu}$ dependence of the Seiberg-Witten is handled in an exact way, then there is an UV/IR mixing phenomenon in the
noncommutative theory defined within the enveloping-algebra formalism. Then, the analysis of the UV/IR mixing was
extended~\cite{Raasakka:2010ev} to  fermions in the fundamental coupled to $U(1)$ gauge fields. Recently, the UV/IR mixing that occurs in the
one-loop propagator of adjoint fermions coupled to U(1) fields and its very interesting implications on neutrino physics has been deeply analysed in refs.~\cite{Horvat:2011iv, Horvat:2011bs, Horvat:2011qn, Horvat:2011qg}.

And yet, up to the best of my knowledge, there is no discussion in the literature on how to obtain systematically
$\theta$-exact  Seiberg-Witten maps for the gauge  of arbitrary gauge groups in an arbitrary unitary representation.
We shall present in this paper such a discussion and compute explicitly the $\theta$-exact Seiberg-Witten map for
an arbitrary gauge field, and  for matter fields as well, up to order three in the number of ordinary gauge fields. This whole contribution is needed to work out --we shall not carry out this lengthy computation in this paper-- the one loop vacuum polarization of the gauge fields, which contains information about the physical phenomena that are consequences of the  UV/IR mixing~\cite{Abel:2006wj}. Let us finally point out that we shall take as starting point of our computations the Seiberg-Witten map
equations that arise when the problem of constructing the Seiberg-Witten map is formulated as a cohomology problem
~\cite{Brace:2001fj, Barnich:2001mc, Barnich:2002pb} --see also ~\cite{Picariello:2001mu}.

The layout of this paper is as follows. In Section 2, we put forward our
procedure for constructing systematically the $\theta$-exact Seiberg-Witten map for the gauge field of an arbitrary gauge group in an arbitrary unitary representation and give the equations that are to be solved by expanding in the gauge coupling constant or in the number of gauge fields. In section 3, we give the
$\theta$-exact Seiberg-Witten map for a general gauge field up to order three  in the number of fields. The construction, by expanding in the gauge coupling constant, of $\theta$-exact Seiberg-Witten maps for matter fields is discussed in Section 4; where also explicit expressions are given up to order three in the number of ordinary gauge fields. In Section 5  we list some pressing open problems which demand the results presented in this paper to be used as a starting point.
In the Appendix we give the very involved value of certain function which gives
the order three --in the number of gauge fields-- contribution to the $\theta$-exact Seiberg-Witten map for matter fields.

\section{The $\theta$-exact Seiberg-Witten map equations and their  solution}

Let the Moyal product, $\star_{h}$, of two functions, $f_1$ and $f_2$, be defined as follows:
\begin{equation*}
(f_1\star_{h}f_2)(x)=\idpq\;{\tilde f}_1(p){\tilde f}_2(q)\; e^{-i\frac{h}{2}(p\wedge q)}\;e^{-i(p+q)x},
\end{equation*}
where $p\wedge q =\theta^{ij}\,p_{i}q_{j}$. ${\tilde f}_1$ and ${\tilde f}_2$ are the Fourier transforms of $f_1$ and $f_2$, respectively.
Let $a_{\mu}$ denote the ordinary gauge field of an arbitrary compact simple gauge group in a arbitrary unitary representation --the case of non-semisimple Lie groups requires a trivial generalization of the
formulae below. Let
 $A_{\mu}[a_\rho;h\theta]$ stand for the noncommutative gauge field which is the image of $a_{\mu}$ under a given Seiberg-Witten map. It has been shown in refs.~\cite{Brace:2001fj, Barnich:2001mc, Barnich:2002pb} that such an  $A_{\mu}$ can be constructed by solving the following problem:
\begin{equation}
\begin{array}{c}
{{\displaystyle \frac{d A_{\mu}}{dh}}=\frac{1}{4}\,\theta^{ij}\{A_{i},\partial_{j}A_{\mu}+F_{j\mu}\}_{\star_{h}}}\\[8pt]
{A_{\mu}[a_\rho;h\theta]\Big{|}_{h=0}=g\,a_{\mu},}
\end{array}
\label{swproblem}
\end{equation}
with $F_{\mu\nu}=\partial_\mu A_\nu-\partial_\nu A_\mu +i[A_\mu,A_\nu]_{\star_{h}}$ and $g$ being the Yang-Mills coupling constant.
Instead of solving the previous problem by expanding in powers of $h\theta^{ij}$, we shall look for a solution to
it that is a power expansion in $g$:
\begin{equation}
A_{\mu}[a_\rho;h\theta]=\sum_{n=1}^{\infty}\;g^n\,A^{(n)}_\mu[a_\rho;h\theta].
\label{ansatzAmu}
\end{equation}
Substituting ~(\ref{ansatzAmu}) in  ~(\ref{swproblem}), one obtains an infinite number of coupled equations which read:
\begin{equation}
\begin{array}{l}
{{\displaystyle\frac{ dA^{(1)}_{\mu}}{dh}}=0, }\\[12pt]
{{\displaystyle\frac{d A^{(2)}_{\mu}}{dh}}=\frac{1}{2}\,\theta^{ij}\{A^{(1)}_{i},\partial_{j}A^{(1)}_{\mu}\}_{\star_{h}}
-\frac{1}{4}\,\theta^{ij}\{A^{(1)}_{i},\partial_{\mu}A^{(1)}_{j}\}_{\star_{h}},  }\\[10pt]
{{\displaystyle\frac{d A^{(3)}_{\mu}}{dh}}=\frac{1}{2}\,\theta^{ij}\{A^{(1)}_{i},\partial_{j}A^{(2)}_{\mu}\}_{\star_{h}}
+\frac{1}{2}\,\theta^{ij}\{A^{(2)}_{i},\partial_{j}A^{(1)}_{\mu}\}_{\star_{h}} }\\[12pt]
{\phantom{\frac{d\phantom{h}}{dh}\,A^{(3)}_{\mu}=}-\frac{1}{4}\,\theta^{ij}\{A^{(2)}_{i},\partial_{\mu}A^{(1)}_{j}\}_{\star_{h}}
-\frac{1}{4}\,\theta^{ij}\{A^{(1)}_{i},\partial_{\mu}A^{(2)}_{j}\}_{\star_{h}} }\\[10pt]
{\phantom{\frac{d\phantom{h}}{dh}\,A^{(3)}_{\mu}=}+\frac{i}{4}\,\theta^{ij}\{A^{(1)}_i,[A^{(1)}_j,A^{(1)}_\mu]_{\star_{h}}\}_{\star_{h}}, }\\[12pt]
{...........}\\[12pt]
{{\displaystyle\frac{d A^{(n)}_{\mu}}{dh}}=\frac{1}{2}\,\theta^{ij}\sum_{m_1+m_2=n}\{A^{(m_1)}_{i},\partial_{j}A^{(m_2)}_{\mu}\}_{\star_{h}}
-\frac{1}{4}\,\theta^{ij}\sum_{m_1+m_2=n}\{A^{(m_1)}_{i},\partial_{\mu}A^{(m_2)}_{j}\}_{\star_{h}}
}\\[8pt]
{\phantom{{\displaystyle\frac{d A^{(n)}_{\mu}}{dh}}=\frac{1}{2}\,\theta^{ij}}
+\frac{i}{4}\,\theta^{ij}\sum_{m_1+m_2+m_3=n}\{A^{(m_1)}_i,[A^{(m_2)}_j,A^{(m_3)}_\mu]_{\star_{h}}\}_{\star_{h}},\;\forall n>3.}
\end{array}
\label{iterativeeqs}
\end{equation}
The initial condition $A_{\mu}[a_\rho;h\theta]\Big{|}_{h=0}=g\,a_{\mu}$ leads to
\begin{equation}
A^{(1)}_{\mu}[a_\rho;h\theta]\Big{|}_{h=0}=a_{\mu}\quad\text{and}\quad A^{(n)}_{\mu}[a_\rho;h\theta]\Big{|}_{h=0}=0,\; \forall n>1.
\label{initialconditions}
\end{equation}
Integrating with respect to $h$ both sides of the equations in ~(\ref{iterativeeqs}) and imposing next the initial conditions in ~(\ref{initialconditions}) , one gets $A^{(n)}_{\mu}[a_{\rho};h\theta]$ in terms of
$A^{(m)}_{\mu}[a_{\rho};h\theta]$, $m=1,...,n-1$:
\begin{equation}
\begin{array}{l}
{A^{(1)}_{\mu}[a_{\rho};h\theta]=a_\mu, \forall h,}\\[12pt]
{A^{(2)}_{\mu}[a_{\rho};h\theta]=\int_{0}^{h}\,dt\,\Big(\frac{1}{2}\,\theta^{ij}\{A^{(1)}_{i},\partial_{j}A^{(1)}_{\mu}\}_{\star_{t}}
-\frac{1}{4}\,\theta^{ij}\{A^{(1)}_{i},\partial_{\mu}A^{(1)}_{j}\}_{\star_{t}}\Big),}\\[12pt]
{A^{(3)}_{\mu}[a_{\rho};h\theta]=\int_{0}^{h}\,dt\,\Big(
\frac{1}{2}\,\theta^{ij}\{A^{(1)}_{i},\partial_{j}A^{(2)}_{\mu}[a_{\rho};t\theta]\}_{\star_{t}}
+\frac{1}{2}\,\theta^{ij}\{A^{(2)}_{i}[a_{\rho};t\theta],\partial_{j}A^{(1)}_{\mu}\}_{\star_{t}} }\\[8pt]
{\phantom{A^{(3)}_{\mu}[a_{\rho};h\theta]=\int_{0}^{h}\,dt\,\Big(}-\frac{1}{4}\,\theta^{ij}\{A^{(2)}_{i}[a_{\rho};t\theta],\partial_{\mu}A^{(1)}_{j}\}_{\star_{t}}
-\frac{1}{4}\,\theta^{ij}\{A^{(1)}_{i},\partial_{\mu}A^{(2)}_{j}[a_{\rho};t\theta]\}_{\star_{t}} }\\[8pt]
{\phantom{A^{(3)}_{\mu}[a_{\rho};h\theta]=\int_{0}^{h}\,dt\,\Big(}+\frac{i}{4}\,\theta^{ij}\{A^{(1)}_i,[A^{(1)}_j,A^{(1)}_\mu]_{\star_{t}}\}_{\star_{t}}, \Big),}\\[12pt]
{...........}\\[12pt]
{A^{(n)}_{\mu}[a_{\rho};h\theta]=\int_{0}^{h}\,dt\,\Big(
\frac{1}{2}\,\theta^{ij}\sum_{m_1+m_2=n}\{A^{(m_1)}_{i},\partial_{j}A^{(m_2)}_{\mu}\}_{\star_{t}}
-\frac{1}{4}\,\theta^{ij}\sum_{m_1+m_2=n}\{A^{(m_1)}_{i},\partial_{\mu}A^{(m_2)}_{j}\}_{\star_{t}}
}\\[12pt]
{\phantom{A^{(n)}_{\mu}[a_{\rho};h\theta]=\int_{0}^{h}\,dt\,\Big(
\frac{1}{2}\,\theta^{ij}}
+\frac{i}{4}\,\theta^{ij}\sum_{m_1+m_2+m_3=n}\{A^{(m_1)}_i,[A^{(m_2)}_j,A^{(m_3)}_\mu]_{\star_{t}}\}_{\star_{t}}\Big),\;\forall n>3.}
\end{array}
\label{iterativeintegration}
\end{equation}
$A^{(m_s)}_\sigma$ is a shorthand for $A^{(m_s)}_\sigma[a_\rho;t\theta]$, whatever $s=1,2,3$ and $\sigma=i,j,\mu$.
The functions displayed in  ~(\ref{iterativeintegration}) give the $\theta$-exact Seiberg-Witten map as an expansion in the Yang-Mills coupling constant or, equivalently, as an expansion in the number of ordinary gauge fields  --$A^{(n)}_{\mu}[a_{\rho};h\theta]$ being the contribution containing $n$ fields $a_\mu$. Closed explicit forms for $A^{(n)}_{\mu}[a_{\rho};h\theta]$, $n=2,3,...$, can be worked out by Fourier transforming the ordinary gauge field. The exponential phase factors characteristic of noncommutative space-time that give rise to UV/IR mixing will thus be clearly exhibited.

Let $G=G_1\times\cdots \times G_N$ be a compact non-semisimple gauge group, $G_i$ being a simple compact group if $i=1,...,s$ and an abelian group if $i=s+1,...,N$. Then,
the results presented in ~(\ref{iterativeintegration}) also hold for $G$, if $g\, a_{\mu}$ is replaced with the whole ordinary gauge field, $v_{\mu}$, defined ~\cite{Brandt:2003fx} as follows
\begin{equation}
v_\mu=\sum_{k=1}^s\,g_k\,(a^{(k)}_\mu)^a\,(T^{(k)})^a\;+\;\sum_{l=s+1}^N\, g_l\,a^{(l)}_\mu\,T^{(l)}.
\label{wholefield}
\end{equation}
$\{(T^{(k)})^a, T^{(l)}\}$, with $a=1,...,\text{dim}\,G_k$ for every $k=1,...,s$, stand for the generators of the gauge group $G$ in a given unitary irreducible representation. Of course, the matrix elements $IJ$ of these generators are always of the form:
\begin{equation*}
(T^{(k)})^a_{IJ}=\delta_{i_1 j_1}\cdots (T^{(k)})^a_{i_k j_k}\cdots\delta_{i_s j_s},\quad  k=1,...,s;\quad
T^{(l)}_{IJ}=\delta_{i_1 j_1}\cdots \delta_{i_s j_s}\,Y^{(l)},\quad l=s+1,...,N.
\end{equation*}

\section{Working out $\theta$-exact closed expressions up to three gauge fields}

In this section we shall show how to use the results in ~(\ref{iterativeintegration}) to generate $\theta$-exact closed expressions in  terms of the gauge fields in momentum space. We think that these expressions will be
most useful in the study, by using Feynman-diagram techniques, of the $\theta$-exact properties of noncommutative field theories defined within the enveloping-algebra formalism.

Let $\{T^a\}$ denote the generators of the compact gauge group in an arbitrary unitary representation. $\{T^{a}\}$ are hermitian and satisfy $[T^a,T^b]=if^{abc} T^c$. Let $a^a_{\mu}(p)\,T^a$ be the Fourier transform of
the ordinary gauge field $a_{\mu}(x)$:
\begin{equation*}
a_{\mu}(x)=\idp\, e^{-ipx}\,a^a_{\mu}(p)T^a.
\end{equation*}
Then, after some computations, one obtains the following $\theta$-exact expression for $A^{(2)}_{\mu}[a_{\rho};h\theta]$  in  ~(\ref{iterativeintegration})
\begin{equation*}
A^{(2)}_{\mu}[a_{\rho};h\theta](x)=\idpp\;e^{-i(p_1+p_2)x}\;\AA_{\mu}^{(2)}[(p_1,\mu_1,a_1),(p_2,\mu_2,a_2);h\theta]\;a^{a_1}_{\mu_1}(p_1)\,a^{a_2}_{\mu_2}(p_2),
\end{equation*}
where
\begin{equation*}
\begin{array}{l}
{\AA_{\mu}^{(2)}[(p_1,\mu_1,a_1),(p_2,\mu_2,a_2);h\theta]=\frac{1}{2}\,\theta^{ij}\,(2 p_{2\,j}\,\delta_{i}^{\mu_1}\delta_{\mu}^{\mu_2}-p_{2\,\mu}\,\delta_{i}^{\mu_1}\delta_{j}^{\mu_2})}\\[4pt]

{\phantom{\AA_{\mu}^{(2)}[(p_1,\mu_1,a_1),(p_2,\mu_2,a_2);h\theta]=} {\displaystyle\big[T^{a_1}T^{a_2}\,\frac{e^{-i\frac{h}{2}p_1\wedge p_2}-1}{p_1\wedge p_2}-T^{a_2}T^{a_1}\,\frac{e^{i\frac{h}{2}p_1\wedge p_2}-1}{p_1\wedge p_2}\big]}.}
\end{array}
\end{equation*}

Let us move on to compute $A^{(3)}_{\mu}[a_{\rho};h\theta]$ in ~(\ref{iterativeintegration}).
  Lengthy computations yield the following $\theta$-exact result:
\begin{equation*}
\begin{array}{l}
{A^{(3)}_{\mu}[a_{\rho};h\theta](x)={\displaystyle\idppp\;e^{-i(p_1+p_2+p_3)x}\;\Big\{}}\\[4pt]
{\phantom{A^{(3)}_{\mu}[a_{\rho};h\theta](x)=}\AA_{\mu}^{(3)}[(p_1,\mu_1,a_1),(p_2,\mu_2,a_2),(p_3,\mu_3,a_3);h\theta]\;a^{a_1}_{\mu_1}(p_1)\,a^{a_2}_{\mu_2}(p_2)\,a^{a_3}_{\mu_3}(p_3)\Big\}},
\end{array}
\end{equation*}
where
\begin{equation*}
\begin{array}{l}
{\AA_{\mu}^{(3)}[(p_1,\mu_1,a_1),(p_2,\mu_2,a_2),(p_3,\mu_3,a_3);h\theta]=}\\[4pt]
{\;\mathbb{P}^{(3)}_{\mu}[(p_1,\mu_1),(p_2,\mu_2),(p_3,\mu_3);\theta]\,\Big(T^{a_1}T^{a_2}T^{a_3}\,
\mathbb{I}(p_1;p_2,p_3;h,\theta)+T^{a_2}T^{a_3}T^{a_1}\,
\mathbb{I}(-p_1;p_2,p_3;h,\theta)\Big)}\\[4pt]
{\quad\quad\quad+\,\mathbb{Q}^{(3)}_{\mu}[\mu_1,\mu_2,\mu_3;\theta]\,\Big(T^{a_1}T^{a_2}T^{a_3}\,
\mathbb{F}(p_1;p_2,p_3;h,\theta)+T^{a_2}T^{a_3}T^{a_1}\,
\mathbb{F}(-p_1;p_2,p_3;h,\theta)\Big)}
\end{array}
\end{equation*}
and
\begin{equation}
\begin{array}{l}
{\mathbb{P}^{(3)}_{\mu}[(p_1,\mu_1),(p_2,\mu_2),(p_3,\mu_3);\theta]=\frac{1}{4}\,\theta^{ij}\theta^{kl}\Big\{
[4(p_{3 l}\,\delta_k^{\mu_2}\delta_i^{\mu_3}\!+\!p_{2 l}\,\delta_i^{\mu_2}\delta_k^{\mu_3})-2(p_3\!-\!p_2)_i\,\delta_k^{\mu_2} \delta_l^{\mu_3}]p_{1j}\,\delta_\mu^{\mu_1}}\\[4pt]
{\phantom{\mathbb{P}^{(3)}_{\mu}[(p_1,\mu_1),(p_2,\mu_2),(p_3,\mu_3)}
+[4(p_{3 l}\,\delta_k^{\mu_2}\delta_\mu^{\mu_3}+p_{2 l}\,\delta_\mu^{\mu_2}\delta_k^{\mu_3})-2(p_3\!-\!p_2)_{\mu}\,\delta_k^{\mu_2} \delta_l^{\mu_3}]\,(p_2\!+\!p_3)_j\,\delta_i^{\mu_1}}\\[4pt]
{\phantom{\mathbb{P}^{(3)}_{\mu}[(p_1,\mu_1),(p_2,\mu_2),(p_3,\mu_3)}
-[2(p_{3 l}\,\delta_k^{\mu_2}\delta_i^{\mu_3}+p_{2 l}\,\delta_i^{\mu_2}\delta_k^{\mu_3})-(p_3\!-\!p_2)_{i}\,\delta_k^{\mu_2} \delta_l^{\mu_3}]\,p_{1 \mu}\,\delta_{j}^{\mu_1}}\\[4pt]
{\phantom{\mathbb{P}^{(3)}_{\mu}[(p_1,\mu_1),(p_2,\mu_2),(p_3,\mu_3)}
-[2(p_{3 l}\,\delta_k^{\mu_2}\delta_j^{\mu_3}+p_{2 l}\,\delta_j^{\mu_2}\delta_k^{\mu_3})-(p_3\!-\!p_2)_{j}\,\delta_k^{\mu_2} \delta_l^{\mu_3}](p_2\!+\!p_3)_\mu\,\delta_{i}^{\mu_1}\Big\},}\\[8pt]
{\mathbb{Q}^{(3)}_{\mu}[\mu_1,\mu_2,\mu_3;\theta]=-\frac{1}{2}\,\theta^{ij}(\delta_i^{\mu_1}\delta_j^{\mu_2}\delta_\mu^{\mu_3}-\delta_i^{\mu_1}\delta_\mu^{\mu_2}\delta_j^{\mu_3}).}
\end{array}
\label{pandq}
\end{equation}
In the previous equation,
\begin{equation*}
\begin{array}{l}
{\mathbb{I}( p_1;p_2,p_3;h,\theta)={\displaystyle\frac{1}{p_2\wedge p_3}\Big[\frac{e^{-i\frac{h}{2}(p_1\wedge p_2+ p_1\wedge p_3+p_2\wedge p_3)}-1}{ p_1\wedge p_2+ p_1\wedge p_3+p_2\wedge p_3}-
\frac{e^{-i\frac{h}{2} p_1\wedge( p_2+p_3)}-1}{p_1\wedge( p_2+p_3)}\Big]},
}\\[12pt]
{\mathbb{F}( p_1;p_2,p_3;h,\theta)={\displaystyle\frac{e^{-i\frac{h}{2}( p_1\wedge p_2+ p_1\wedge p_3+p_2\wedge p_3)}-1}{ p_1\wedge p_2+p_1\wedge p_3+p_2\wedge p_3}},}\\[12pt]
{\mathbb{I}(-p_1;p_2,p_3;h,\theta)=\mathbb{I}( p_1;p_2,p_3;h,\theta)|_{p_1\rightarrow -p_1},\quad\mathbb{F}(-p_1;p_2,p_3;h,\theta)=\mathbb{F}( p_1;p_2,p_3;h,\theta)|_{p_1\rightarrow -p_1}.}
\end{array}
\end{equation*}

\section{Computing the $\theta$-exact Seiberg-Witten maps for matter fields}

The method that we have presented in the previous sections can be successfully applied to the case of
matter fields. We shall deal with the  non-hybrid (i.e., standard) Seibeg-Witten maps only, but it will become clear that
the procedure can also be applied to the hybrid~\cite{Schupp:2001we} case.

Let $\Psi$ denote a noncommutative matter field (either a boson or a fermion) which changes under
a noncommutative BRS transformation, $s_{NC}$, characterised by the noncommutative ghost $\Lambda$ as follows: $s_{NC}\,\Psi=-i\Lambda\star_{h}\Psi$. Let $\psi$ denote the ordinary matter field whose image under the Seiberg-Witten map is
$\Psi$. Under the ordinary BRS operator, $s$, $\psi$ transforms thus $s\psi=-i\lambda\,\psi$, where $\lambda=\lambda^a T^a$.
$\{T^a\}$ are the generators of an arbitrary compact Lie group in an arbitrary unitary representation. Then, as shown in refs.~\cite{ Barnich:2002pb}, a non-hybrid Seiberg-Witten map from $\psi$ to $\Psi=\Psi[a_\rho,\psi;h\theta]$ can be obtained by solving the problem:
\begin{equation}
\begin{array}{c}
{{\displaystyle \frac{d\Psi}{dh}=\frac{1}{2}\,\theta^{ij}\,A_i\star_{h}\partial_j\Psi+\frac{i}{4}\,\theta^{ij}\,A_i\star_{h} A_j\star_{h}\Psi}}\\[8pt]
{\Psi[a_\rho,\psi;h\theta]\Big{|}_{h=0}=\psi}.
\end{array}
\label{psiproblem}
\end{equation}
Let us consider for the time being that the ordinary gauge group is simple and  introduce the following expansion of $\Psi[a_\rho,\psi;h\theta]$ in terms of the  gauge coupling constant $g$:
\begin{equation}
\Psi[a_\rho,\psi;h\theta]=\sum_{n=0}^{\infty}\,g^n\,\Psi[a_\rho,\psi;h\theta].
\label{psiexpan}
\end{equation}
Then substituting ~(\ref{ansatzAmu}) and ~(\ref{psiexpan}) in the  differential equation in ~(\ref{psiproblem}), one
obtains
\begin{equation}
\begin{array}{l}
{{\displaystyle\frac{ d\Psi^{(0)}}{dh}}=0, }\\[12pt]
{{\displaystyle\frac{d \Psi^{(1)}}{dh}}=\frac{1}{2}\,\theta^{ij}A^{(1)}_{i}\star_{h}\partial_{j}\Psi^{(0)}, }\\[12pt]
{{\displaystyle\frac{d \Psi^{(2)}}{dh}}=\frac{1}{2}\,\theta^{ij}A^{(1)}_{i}\star_{h}\partial_{j}\Psi^{(1)}+
\frac{1}{2}\,\theta^{ij}A^{(2)}_{i}\star_{h}\partial_{j}\Psi^{(0)}+\frac{i}{4}\,\theta^{ij}\,A^{(1)}_i\star_h A^{(1)}_j\star_h \Psi^{(0)},}\\[12pt]
{{\displaystyle\frac{d \Psi^{(3)}}{dh}}=\frac{1}{2}\,\theta^{ij}A^{(3)}_{i}\star_{h}\partial_{j}\Psi^{(0)}+
\frac{1}{2}\,\theta^{ij}A^{(2)}_{i}\star_{h}\partial_{j}\Psi^{(1)}+ \frac{1}{2}\,\theta^{ij}A^{(1)}_{i}\star_{h}\partial_{j}\Psi^{(2)} }\\[10pt]
{\phantom{{\displaystyle\frac{d \Psi^{(3)}}{dh}}=}+\frac{i}{4}\,\theta^{ij}\,A^{(2)}_i\star_h A^{(1)}_j\star_h \Psi^{(0)}+\frac{i}{4}\,\theta^{ij}\,A^{(1)}_i\star_h A^{(2)}_j\star_h \Psi^{(0)}+\frac{i}{4}\,\theta^{ij}\,A^{(1)}_i\star_h A^{(1)}_j\star_h \Psi^{(1)},}\\[12pt]
{...........}\\[12pt]
{{\displaystyle\frac{d \Psi^{(n)}}{dh}}=\frac{1}{2}\,\theta^{ij}\sum_{m_1+m_2=n}A^{(m_1)}_{i}\star_{h}\partial_{j}\Psi^{(m_2)}
+\frac{i}{4}\,\theta^{ij}\sum_{m_1+m_2+m_3=n} A^{(m_1)}_i\star_{h}A^{(m_2)}_j\star_{h}\Psi^{(m_3)},\; \forall n>3.}
\end{array}
\label{iterativepsieqs}
\end{equation}
The initial condition $\Psi[a_\rho,\psi;h\theta]\Big{|}_{h=0}=\psi$ leads to
\begin{equation}
\Psi^{(0)}_{\mu}[a_\rho;h\theta]\Big{|}_{h=0}=\psi\quad\text{and}\quad \Psi^{(n)}_{\mu}[a_\rho,\psi;h\theta]\Big{|}_{h=0}=0,\; \forall n>0.
\label{iteratinitial}
\end{equation}
Integration over $h$ of both sides of ~(\ref{iterativepsieqs}) and the initial conditions in ~(\ref{iteratinitial})
yield the following solution  to ~(\ref{psiproblem}):
\begin{equation}
\begin{array}{l}
{ \Psi^{(0)}[a_\rho,\psi;h\theta]=\psi, }\\[12pt]
{\Psi^{(1)}[a_\rho,\psi;h\theta]=\int_{0}^{h}\,dt\,\Big(\frac{1}{2}\,\theta^{ij}A^{(1)}_{i}\star_{t}\partial_{j}\Psi^{(0)}\Big), }\\[12pt]
{\Psi^{(2)}[a_\rho,\psi;h\theta]=\int_{0}^{h}\,dt\,\Big(\frac{1}{2}\,\theta^{ij}A^{(1)}_{i}\star_{t}
\partial_{j}\Psi^{(1)}[t\theta]+
\frac{1}{2}\,\theta^{ij}\, A^{(2)}_i[t\theta]\star_t\partial_{j}\Psi^{(0)}+\frac{i}{4}\,\theta^{ij}\,A^{(1)}_i\star_t A^{(1)}_j\star_t \Psi^{(0)}\Big),}\\[12pt]
{\Psi^{(3)}[a_\rho,\psi;h\theta]=\int_{0}^{h}\,dt\,\Big(
\frac{1}{2}\,\theta^{ij}A^{(3)}_{i}[t\theta]\star_{t}\partial_{j}\Psi^{(0)}+
\frac{1}{2}\,\theta^{ij}A^{(2)}_{i}[t\theta]\star_{t}\partial_{j}\Psi^{(1)}[t\theta]}\\[10pt]
{\phantom{\Psi^{(3)}[a_\rho,\psi;h\theta]=\int_{0}^{h}\,dt\,\Big(
\frac{1}{2}\,\theta^{ij}A^{(3)}_{i}[t\theta]\star_{t}\partial_{j}\Psi^{(0)}+
\frac{1}{2}\,\theta^{ij}A^{(2)}_{i}[t\theta]\star_{t} \partial_{j} \Psi              }
+\frac{1}{2}\,\theta^{ij}A^{(1)}_{i}\star_{t}\partial_{j}\Psi^{(2)}[t\theta] }\\[10pt]
{\quad\quad\quad\quad +\frac{i}{4}\,\theta^{ij}\,A^{(2)}_i[t\theta]\star_t A^{(1)}_j\star_t \Psi^{(0)}+\frac{i}{4}\,\theta^{ij}\,A^{(1)}_i\star_t A^{(2)}_j[t\theta]\star_t \Psi^{(0)}+\frac{i}{4}\,\theta^{ij}\,A^{(1)}_i\star_t A^{(1)}_j\star_t \Psi^{(1)}[t\theta]\underline{}\Big),}\\[12pt]
{...........}\\[12pt]
{\Psi^{(n)}[a_\rho,\psi;h\theta]=\int_{0}^{h}\,dt\,\Big(\frac{1}{2}\,\theta^{ij}
\sum_{m_1+m_2=n}A^{(m_1)}_{i}[t\theta]\star_{t}\partial_{j}\Psi^{(m_2)}[t\theta]}\\[10pt]
{\phantom{\Psi^{(n)}[a_\rho,\psi;h\theta]=\int_{0}^{h}\,dt\,\Big(\frac{1}{2}\,\theta^{ij}}
+\frac{i}{4}\,\theta^{ij}\sum_{m_1+m_2+m_3=n} A^{(m_1)}_i[t\theta]\star_{t}A^{(m_2)}_j[t\theta]\star_{t}\Psi^{(m_3)}[t\theta]\Big),\;\forall n>3.}
\end{array}
\label{iterativesolpsi}
\end{equation}
$A^{(m)}_\mu$ is given in ~(\ref{iterativeintegration}).

We have worked out $\Psi^{(1)}[a_\rho,\psi;h\theta]$,  $\Psi^{(2)}[a_\rho,\psi;h\theta]$ and
$\Psi^{(3)}[a_\rho,\psi;h\theta]$
in terms of the Fourier transforms of the ordinary fields $a_\mu$ and $\psi$. Our results run thus:
\begin{equation}
\begin{array}{l}
{\Psi^{(1)}_{A}[a_\rho,\psi;h\theta](x)=\idpp\;e^{-i(p_1+p_2)x}\;
\big(\MM^{(1)}[(p_1,\mu_1,a_1),p_2;h\theta]\big)_A^{B}\;a^{a_1}_{\mu_1}(p_1)\psi_B(p_2),}\\[12pt]
{\Psi^{(2)}_{A}[a_\rho,\psi;h\theta](x)={\displaystyle\idppp\;e^{-i(p_1+p_2+p_3)x}\;\Big\{}}\\[10pt]
{\phantom{A^{(3)}_{\mu}[a_{\rho};h\theta](x)=}\big(\MM^{(2)}[(p_1,\mu_1,a_1),(p_2,\mu_2,a_2),p_3;h\theta]\big)_A^B\;
a^{a_1}_{\mu_1}(p_1)\,a^{a_2}_{\mu_2}(p_2)\,\psi_B(p_3)\Big\},}\\[12pt]
{\Psi^{(3)}_{A}[a_\rho,\psi;h\theta](x)={\displaystyle\idpppp\;e^{-i(p_1+p_2+p_3+p_4)x}\;\Big\{}}\\[10pt]
{\phantom{A^{(3)}_{\mu}[a_{\rho}\quad}\big(\MM^{(3)}[(p_1,\mu_1,a_1),(p_2,\mu_2,a_2),(p_3,\mu_3,a_3),p_4;h\theta]\big)_A^B\;
a^{a_1}_{\mu_1}(p_1)\,a^{a_2}_{\mu_2}(p_2)\,a^{a_3}_{\mu_3}(p_3)\,\psi_B(p_4)\Big\}},
\end{array}
\label{sundrypsis}
\end{equation}
where
\begin{equation*}
\begin{array}{l}
{\big(\MM^{(1)}[(p_1,\mu_1,a_1),p_2;h\theta]\big)_A^{B}=\big(T^{a_1}\big)_A^B\;\theta^{ij}\delta_i^{\mu_1}p_{2 j}\;
{\displaystyle\frac{e^{-i\frac{h}{2}p_1\wedge p_2}-1}{p_1\wedge p_2}},}\\[12pt]
{\big(\MM^{(2)}[(p_1,\mu_1,a_1),(p_2,\mu_2,a_2),p_3;h\theta]\big)_A^B=\big(T^{a_1}T^{a_2}\big)_A^B\,\Big\{}\\[10pt]
{\quad\quad\quad\theta^{ij}\theta^{kl}\,\delta_i^{\mu_1}\delta_k^{\mu_2}(p_2\!+\!p_3)_j\, p_{3 l}\,
{\displaystyle\frac{1}{p_2\wedge p_3}\Big[\frac{e^{-i\frac{h}{2}(p_1\wedge p_2+ p_1\wedge p_3+p_2\wedge p_3)}-1}{ p_1\wedge p_2+ p_1\wedge p_3+p_2\wedge p_3}-
\frac{e^{-i\frac{h}{2} p_1\wedge( p_2+p_3)}-1}{p_1\wedge( p_2+p_3)}\Big]}}\\[10pt]
{\phantom{\quad\quad\quad\quad\theta^{ij}\theta^{kl}}+\frac{1}{2}\,\theta^{ij}\theta^{kl}\;
[2(p_{2 l}\,\delta_k^{\mu_1}\delta_i^{\mu_2}+p_{1 l}\,\delta_k^{\mu_2}\delta_i^{\mu_1})-(p_2\!-\!p_1)_{i}\,\delta_k^{\mu_1} \delta_l^{\mu_2}]\,p_{3 j}}\\[10pt]
{\phantom{\quad\quad\quad\quad\theta^{ij}\theta^{kl}+\frac{1}{2}\,\theta^{ij}\theta^{kl}}{\displaystyle\frac{1}{p_1\wedge p_2}\Big[\frac{e^{-i\frac{h}{2}(p_1\wedge p_2+ p_1\wedge p_3+p_2\wedge p_3)}-1}{ p_1\wedge p_2+ p_1\wedge p_3+p_2\wedge p_3}-
\frac{e^{-i\frac{h}{2}( p_1+p_2)\wedge p_3-1}}{( p_1+p_2)\wedge p_3}\Big]}}\\[10pt]
{\phantom{\phantom{\theta^{kl}+\frac{1}{2}\,\theta^{ij}\theta^{kl}}{\displaystyle\frac{1}{p_1\wedge p_2}\Big[\frac{e^{-i\frac{h}{2}(p_1\wedge p_2+ p_1\wedge p_3+p_2\wedge p_3)}-1}{ p_1\wedge p_2+ p_1\wedge p_3+p_2\wedge p_3}}} -\frac{1}{2}\,\theta^{ij}\;\delta_i^{\mu_1}\delta_j^{\mu_2}\,
{\displaystyle\Big[\frac{e^{-i\frac{h}{2}(p_1\wedge p_2+ p_1\wedge p_3+p_2\wedge p_3)}-1}{ p_1\wedge p_2+ p_1\wedge p_3+p_2\wedge p_3}\Big]}\Big\}.}
\end{array}
\end{equation*}
$\MM^{(3)}[(p_1,\mu_1,a_1),(p_2,\mu_2,a_2),(p_3,\mu_3,a_3),p_4;h\theta]\big)_A^B$ in ~(\ref{sundrypsis}) is defined by a very involved expression that we have committed to the Appendix.

As in the gauge field case, let $G=G_1\times\cdots G_N$ be a compact non-semisimple gauge group, $G_i$ being a simple compact group if $i=1,...,s$ and an abelian group if $i=s+1,...,N$. Then,
the results presented in ~(\ref{iterativesolpsi}) and ~(\ref{sundrypsis}) also hold for $G$, if $g\, a_{\mu}$ is replaced with the whole ordinary gauge field, $v_{\mu}$, as defined in ~(\ref{wholefield}).

Finally, if we have a matter field, $\Upsilon$, transforming under the adjoint, ie,
\begin{equation*}
s_{NC}\,\Upsilon\,=\,-i\,[\Lambda,\Upsilon]_{\star_h},
\end{equation*}
one may use the equation
\begin{equation}
{\displaystyle \frac{d\Upsilon}{dh}}=\frac{1}{2}\,\theta^{ij}\,\{A_i,\partial_j\Upsilon\}_{\star_h}+
\frac{i}{4}\,\theta^{ij}\,\{A_i,[ A_j,\Upsilon]_{\star_h}\}_{\star_h}
\label{adjointequ}
\end{equation}
to obtain an $\theta$-exact Seiberg-Witten map giving $\Upsilon$, as done for the matter field $\Psi$ above. Notice that the equation in
~(\ref{adjointequ}) can be obtained from  ~(\ref{swproblem}) by setting $A_0=\Upsilon$ and considering  time independent fields in a five dimensional space-time with space-space noncommutativity only. We have worked out up to order two in the number of ordinary gauge fields the $\theta$-exact Seiberg-Witten map that is a solution to the equation  in ~(\ref{adjointequ}). The results that one obtains are far more involved that the ones we have already given here, so we shall report on them elsewhere.

\section{Future directions}

Furnished with the expressions given in Section 3 and 4, one may compute the one-loop propagators  of gauge fields and matter fields for arbitrary compact gauge groups in arbitrary unitary representations and
obtain the precise momentum structure of the noncommutative UV/IR mixing. This is, of course, a very lengthy computation, and this is why we have not tackled it here. Further, the results presented in section 4 will be needed to work out the gauge anomaly equations when the noncommutative gauge theory is defined by using $\theta$-exact Seiberg-Witten maps. Notice that the calculations and discussion given in refs.~\cite{Martin:2002nr, Brandt:2003fx}  rest on the Seiberg-Witten map being defined as a formal power series in $\theta^{\mu\nu}$ and, certainly, the UV/IR behaviour of the theories so defined is not the same as the UV/IR behaviour of the corresponding
theories defined by means of  $\theta$-exact Seiberg-Witten maps. Hence, it is a must to repeat the analysis and computations carried out in refs.~\cite{Martin:2002nr, Brandt:2003fx} when $\theta$-exact Seiberg-Witten
maps are used. Finally, after obtaining by using cohomological techniques an equation analogous to ~(\ref{psiproblem}) for the case of hybrid Seiberg-Witten maps, it would be advisable to work out explicit expressions
like the ones computed in this paper. Of course, it is not difficult to make an educated guess and state that a hybrid --see ~\cite{Schupp:2001we}, for its definition-- Seiberg-Witten map is furnished by the solution to
\begin{equation*}
\begin{array}{l}
{{\displaystyle \frac{d\Phi}{dh}}=\frac{1}{2}\,\theta^{ij}\,A_i\star_{h}\partial_j\Phi+\frac{i}{4}\,\theta^{ij}\,A_i\star_{h} A_j\star_{h}\Phi}\\[8pt]
{\phantom{{\displaystyle \frac{d\Phi}{dh}}=}
+\frac{1}{2}\,\theta^{ij}\,\partial_j\Phi\star_{h} B_i- \frac{i}{4}\,\theta^{ij}\,\Phi\star_{h} B_j\star_{h} B_i-\frac{i}{2}\,\theta^{ij}\,A_i\star_{h}\Phi\star_{h} B_j}\\[8pt]
{\Phi[a_\rho,b_\rho,\phi;h\theta]\Big{|}_{h=0}=\phi},
\end{array}
\end{equation*}
where $A_i$ and $B_i$ denote, respectively, the gauge fields acting from the left and from the right on the matter field $\Phi$. Under noncommutative BRS transformations $\Phi$ would transform as follows:
\begin{equation*}
s_{NC}\,\Phi\,=\,-i\,\Lambda\star_{h}\Phi\,+\,i\,\Phi\star_{h}\Omega,\quad s^2_{NC}\Phi=0,
\end{equation*}
where $s_{NC}\,A_\mu=\partial_\mu\Lambda+i[A_\mu,\Lambda]_{\star_h}$, $s_{NC}\,B_\mu=\partial_\mu\Omega+i[B_\mu,\Omega]_{\star_h}$, $s_{NC}\,\Lambda=-i\,\Lambda\star_h\Lambda$ and
$s_{NC}\,\Omega=-i\,\Omega\star_h\Omega$.

\newpage
\section{Appendix}

Is this Appendix we define $\MM^{(3)}[(p_1,\mu_1,a_1),(p_2,\mu_2,a_2),(p_3,\mu_3,a_3),p_4;h\theta]\big)_A^B$ in ~(\ref{sundrypsis}). We shall first introduce a number of momentum dependent functions:
\begin{equation}
\begin{array}{l}
{\Sigma(p_1,p_2,p_3,p_4,\theta)=\sum_{i<j}\,p_1\wedge p_j=(p_1+p_2+p_3)\wedge p_4+p_2\wedge p_3 + p_1\wedge (p_2+p_3),}\\[10pt]
{\Theta(p_1,p_2,p_3,p_4,\theta)=(p_1+p_2+p_3)\wedge p_4+p_2\wedge p_3 - p_1\wedge (p_2+p_3),}\\[10pt]
{\mathbb{L}_1(p_1,p_2,p_3,p_4;h,\theta)=}\\[4pt]
{\quad\displaystyle{
\frac{1}{p_1\wedge(p_2+p_3)+p_2\wedge p_3}\Big[
\frac{e^{-i\frac{h}{2}\Sigma(p_1,p_2,p_3,p_4,\theta)}-1}{\Sigma(p_1,p_2,p_3,p_4,\theta)}-\frac{e^{-i\frac{h}{2}(p_1+p_2+p_3)\wedge p_4}-1}{(p_1+p_2+p_3)\wedge p_4}\Big],}}\\[18pt]
{\mathbb{L}_2(p_1,p_2,p_3,p_4;h,\theta)=}\\[4pt]
{\quad\displaystyle{\frac{1}{-p_1\wedge(p_2+p_3)+p_2\wedge p_3}\Big[
\frac{e^{-i\frac{h}{2}\Theta(p_1,p_2,p_3,p_4,\theta)}-1}{\Theta(p_1,p_2,p_3,p_4,\theta)}-\frac{e^{-i\frac{h}{2}(p_1+p_2+p_3)\wedge p_4}-1}{(p_1+p_2+p_3)\wedge p_4}\Big],}}\\[18pt]
{\mathbb{K}_1(p_1,p_2,p_3,p_4;h,\theta)=\displaystyle{
\frac{1}{p_2\wedge p_3}\Big\{\;\mathbb{L}_1(p_1,p_2,p_3;h,\theta)}}\\[8pt]
{\quad\displaystyle{-\frac{1}{p_1\wedge (p_2+p_3)}
\Big[\frac{e^{-i\frac{h}{2}[(p_1+p_2+p_3)\wedge p_4+p_1\wedge(p_2+p_3)]}-1}{(p_1+p_2+p_3)\wedge p_4+p_1\wedge(p_2+p_3)}-\frac{e^{-i\frac{h}{2}(p_1+p_2+p_3)\wedge p_4}-1}{(p_1+p_2+p_3)\wedge p_4}\Big]\Big\}},}\\[18pt]
{\mathbb{K}_2(p_1,p_2,p_3,p_4;h,\theta)=
\displaystyle{\frac{1}{p_2\wedge p_3}\Big\{\,\mathbb{L}_2(p_1,p_2,p_3;h,\theta) }}\\[8pt]
{\quad\displaystyle{+\frac{1}{p_1\wedge (p_2+p_3)}
\Big[\frac{e^{-i\frac{h}{2}[(p_1+p_2+p_3)\wedge p_4-p_1\wedge(p_2+p_3)]}-1}{(p_1+p_2+p_3)\wedge p_4-p_1\wedge(p_2+p_3)}-\frac{e^{-i\frac{h}{2}(p_1+p_2+p_3)\wedge p_4}-1}{(p_1+p_2+p_3)\wedge p_4}\Big]\Big\}},}\\[18pt]
{\mathbb{K}_3(p_1,p_2,p_3,p_4;h,\theta)=}\\[4pt]
{\quad\displaystyle{\frac{1}{(p_1\wedge p_2)(p_3\wedge p_4)} \Big[
\frac{e^{-i\frac{h}{2}\Sigma(p_1,p_2,p_3,p_4,\theta)}-1}{\Sigma(p_1,p_2,p_3,p_4,\theta)}-
\frac{e^{-i\frac{h}{2}[p_1\wedge p_2+(p_1+p_2)\wedge(p_3+ p_4)]}-1}{p_1\wedge p_2+(p_1+p_2)\wedge(p_3+ p_4)}}}\\[8pt]
{\quad\quad\quad\quad\quad\quad\quad\quad\quad\quad\displaystyle{-\frac{e^{-i\frac{h}{2}[p_3\wedge p_4+(p_1+p_2)\wedge(p_3+ p_4)]}-1}{p_3\wedge p_4+(p_1+p_2)\wedge(p_3+ p_4)}+
\frac{e^{-i\frac{h}{2} (p_1+p_2)\wedge(p_3+ p_4)}-1}{(p_1+p_2)\wedge(p_3+ p_4)}
\Big],}}\\[18pt]
\end{array}
\label{lsandks}
\end{equation}

\begin{equation}
\begin{array}{l}
{\mathbb{K}_4(p_1,p_2,p_3, p_4;h,\theta)=}\\[4pt]
{\quad\displaystyle{
\frac{1}{p_3\wedge p_4}\Big\{
\frac{1}{p_2\wedge p_3+(p_2+p_3)\wedge p_4}\Big[
\frac{e^{-i\frac{h}{2}\Sigma(p_1,p_2,p_3,p_4,\theta)}-1}{\Sigma(p_1,p_2,p_3,p_4,\theta)}-\frac{e^{-i\frac{h}{2}p_1\wedge( p_2+p_3+ p_4)}-1}{p_1\wedge (p_2+p_3 + p_4)}\Big]}}\\[10pt]
{\quad\displaystyle{-\frac{1}{p_2\wedge (p_3+p_4)}
\Big[\frac{e^{-i\frac{h}{2}[p_1\wedge (p_2+p_3 + p_4)+p_2\wedge(p_3+p_4)]}-1}{p_1\wedge (p_2+p_3 + p_4)+p_2\wedge(p_3+p_4)}-\frac{e^{-i\frac{h}{2}p_1\wedge ( p_2+p_3 + p_4)}-1}{p_1\wedge ( p_2+p_3 + p_4)}\Big]\Big\}},}\\[18pt]
{\mathbb{K}_5(p_1,p_2,p_3, p_4;h,\theta)=}\\[4pt]
{\quad\displaystyle{
\frac{1}{p_2\wedge p_3}\Big\{
\frac{1}{p_2\wedge p_3+(p_2+p_3)\wedge p_4}\Big[
\frac{e^{-i\frac{h}{2}\Sigma(p_1,p_2,p_3,p_4,\theta)}-1}{\Sigma(p_1,p_2,p_3,p_4,\theta)}-\frac{e^{-i\frac{h}{2}p_1\wedge( p_2+p_3+ p_4)}-1}{p_1\wedge (p_2+p_3 + p_4)}\Big]}}\\[10pt]
{\quad\displaystyle{-\frac{1}{(p_2+p_3)\wedge p_4}
\Big[\frac{e^{-i\frac{h}{2}[p_1\wedge (p_2+p_3 + p_4)+(p_2+p_3)\wedge p_4]}-1}{p_1\wedge (p_2+p_3 + p_4)+(p_2+p_3)\wedge p_4}-\frac{e^{-i\frac{h}{2}p_1\wedge ( p_2+p_3 + p_4)}-1}{p_1\wedge ( p_2+p_3 + p_4)}\Big]\Big\}},}\\[18pt]
{\mathbb{K}_6(p_1,p_2,p_3, p_4;h,\theta)=
\displaystyle{\frac{1}{p_2\wedge p_3+(p_2+p_3)\wedge p_4}\Big[
\frac{e^{-i\frac{h}{2}\Sigma(p_1,p_2,p_3,p_4,\theta)}-1}{\Sigma(p_1,p_2,p_3,p_4,\theta)}-\frac{e^{-i\frac{h}{2}p_1\wedge( p_2+p_3+ p_4)}-1}{p_1\wedge (p_2+p_3 + p_4)}}\Big]}\\[18pt]
{\mathbb{K}_7(p_1,p_2,p_3,p_4;h,\theta)=\displaystyle{
\frac{1}{p_1\wedge p_2}\Big[
\frac{e^{-i\frac{h}{2}\Sigma(p_1,p_2,p_3,p_4,\theta)}-1}{\Sigma(p_1,p_2,p_3,p_4,\theta)}-\frac{e^{-i\frac{h}{2}[(p_1+p_2)\wedge (p_3+p_4)+p_3\wedge p_4]}-1}{(p_1+p_2)\wedge(p_3+p_4)+p_3\wedge p_4}\Big],}}\\[18pt]
{\mathbb{K}_8(p_1,p_2,p_3,p_4;h,\theta)=\displaystyle{
\frac{1}{p_2\wedge p_3}\Big[
\frac{e^{-i\frac{h}{2}\Sigma(p_1,p_2,p_3,p_4,\theta)}-1}{\Sigma(p_1,p_2,p_3,p_4,\theta)}-\frac{e^{-i\frac{h}{2}[p_1\wedge (p_2+p_3+p_4)+(p_2+p_3)\wedge p_4]}-1}{p_1\wedge(p_2+p_3+p_4)+(p_2+p_3)\wedge p_4}\Big],}}\\[18pt]
{\mathbb{K}_9(p_1,p_2,p_3,p_4;h,\theta)=\displaystyle{
\frac{1}{p_3\wedge p_4}\Big[
\frac{e^{-i\frac{h}{2}\Sigma(p_1,p_2,p_3,p_4,\theta)}-1}{\Sigma(p_1,p_2,p_3,p_4,\theta)}-\frac{e^{-i\frac{h}{2}[p_1\wedge (p_2+p_3+p_4)+p_2\wedge (p_3 + p_4)]}-1}{p_1\wedge(p_2+p_3+p_4)+ p_2\wedge( p_3 + p_4)}\Big].}}
\end{array}
\label{plentyofks}
\end{equation}

Bearing in mind the definitions in ~(\ref{pandq}), ~(\ref{lsandks}) and ~(\ref{plentyofks}), we are finally ready to give $\MM^{(3)}[(p_1,\mu_1,a_1),(p_2,\mu_2,a_2),(p_3,\mu_3,a_3),p_4;h\theta]\big)_A^B$ in ~(\ref{sundrypsis}):

\begin{equation*}
\begin{array}{l}
{\big(\MM^{(3)}[(p_1,\mu_1,a_1),(p_2,\mu_2,a_2),(p_3,\mu_3,a_3),p_4;h\theta]\big)_A^B=(T^{a_1}T^{a_2}T^{a_3})_A^B\Big\{}\\[8pt]
{\quad\quad\theta^{mn}\,p_{4 n}\;\Big[\mathbb{P}^{(3)}_{m}[(p_1,\mu_1),(p_2,\mu_2),(p_3,\mu_3);\theta]\,
\mathbb{K}_1(p_1,p_2,p_3,p_4;h,\theta)}\\[8pt]
{\phantom{\big(\MM^{(3)}[(p_1,\mu_1,a_1),(p_2,\mu_2,a_2),(p_3,\mu_3,a_3),p_4}
 +\mathbb{Q}^{(3)}_{m}[\mu_1,\mu_2,\mu_3;\theta]\,
\mathbb{L}_1(p_1,p_2,p_3,p_4;h,\theta)}\\[8pt]
{\phantom{\quad\quad\theta^{mn}\,p_{4 n}\;\Big[}
+\mathbb{P}^{(3)}_{m}[(p_3,\mu_3),(p_1,\mu_1),(p_2,\mu_2);\theta]\,
\mathbb{K}_2(p_3,p_1,p_2,p_4;h,\theta)}\\[8pt]
{\phantom{\;\big(\MM^{(3)}[(p_1,\mu_1,a_1),(p_2,\mu_2,a_2),(p_3,\mu_3,a_3),p_4}
+\mathbb{Q}^{(3)}_{m}[\mu_3,\mu_1,\mu_2;\theta]\,
\mathbb{L}_2(p_3,p_1,p_2,p_4;h,\theta)\Big]}\\[8pt]
{\quad\quad +\theta^{ij}\theta^{mn}\theta^{kl}\;
\Big[ }\\[8pt]
{\quad\frac{1}{2}\,(p_3\!+\!p_4)_j
[2(p_{2 l}\,\delta_k^{\mu_1}\delta_i^{\mu_2}+p_{1 l}\,\delta_k^{\mu_2}\delta_i^{\mu_1})-(p_2\!-\!p_1)_{i}\,\delta_k^{\mu_1} \delta_l^{\mu_2}]\,\delta_{m}^{\mu_3}\,p_{4n}
\;\mathbb{K}_3(p_1,p_2,p_3,p_4;h,\theta) }\\[8pt]
{\quad\quad\quad\quad\quad +\delta_i^{\mu_1}\delta_{m}^{\mu_2}\delta_k^{\mu_3}\,(p_2\!+\!p_3\!+\!p_4)_j\,(p_3\!+\!p_4)_n\,p_{4l}\,
\;\mathbb{K}_4(p_1,p_2,p_3,p_4;h,\theta)}\\[8pt]
{\;+\frac{1}{2}\,\delta_{i}^{\mu_1}(p_2\!+\!p_3\!+\!p_4)_{j} p_{4n}\,
[2(p_{3 l}\,\delta_k^{\mu_2}\delta_m^{\mu_3}+p_{2 l}\,\delta_m^{\mu_2}\delta_k^{\mu_3})-(p_3\!-\!p_2)_{m}\,\delta_k^{\mu_2} \delta_l^{\mu_3}]
\;\mathbb{K}_5(p_1,p_2,p_3,p_4;h,\theta)\Big] }\\[8pt]
{\quad\quad\quad\quad\quad-\frac{1}{2}\,\theta^{ij}\theta^{kl}\,\delta_i^{\mu_1}\delta_{k}^{\mu_2}\delta_l^{\mu_3}\,(p_2\!+\!p_3\!+\!p_4)_j\;
\mathbb{K}_6(p_1,p_2,p_3,p_4;h,\theta)}\\[8pt]
{\quad\quad -\frac{1}{4}\,\theta^{ij}\theta^{kl}\Big[
[2(p_{2 l}\,\delta_k^{\mu_1}\delta_i^{\mu_2}+p_{1 l}\,\delta_k^{\mu_2}\delta_i^{\mu_1})-(p_2\!-\!p_1)_{i}\,\delta_k^{\mu_1} \delta_l^{\mu_2}]\,\delta_{j}^{\mu_3}\,
\mathbb{K}_7(p_1,p_2,p_3,p_4;h,\theta)}\\[8pt]
{\phantom{\quad\quad -\frac{1}{4}\,\theta^{ij}\theta^{kl}\Big[}
+\delta_i^{\mu_1}\,
[2(p_{3 l}\,\delta_k^{\mu_2}\delta_j^{\mu_3}+p_{2 l}\,\delta_j^{\mu_2}\delta_k^{\mu_3})-(p_3\!-\!p_2)_{j}\,\delta_k^{\mu_2} \delta_l^{\mu_3}]
\;\mathbb{K}_8(p_1,p_2,p_3,p_4;h,\theta) }\\[8pt]
{\phantom{\quad\quad -\frac{1}{4}\,\theta^{ij}\theta^{kl}\Big[}
+\,2\,\delta_i^{\mu_1}\,\delta_j^{\mu_2}\,\delta_k^{\mu_3}\,p_{4l}\;\mathbb{K}_9(p_1,p_2,p_3,p_4;h,\theta)\Big]\Big\}.}
\end{array}
\end{equation*}

\section{Acknowledgements}
I thank  P. Schupp for very useful comments on the content of this paper.
This work has been financially supported in part by MICINN through grant
FPA2008-04906.

\end{document}